\documentclass[11pt,journal]{IEEEtran}
\usepackage{cite}
\usepackage[pdftex]{graphicx}
\DeclareGraphicsExtensions{.pdf,.jpeg,.png}
\usepackage[cmex10]{amsmath}
\usepackage{array}
\usepackage{mdwmath}
\usepackage{mdwtab}
\usepackage{fixltx2e}

\usepackage{subfig}
\usepackage{fixltx2e}
\usepackage{stfloats}
\begin{document}

\title{Cosmic Microwave Background Images}


 \author{Diego~Herranz\IEEEmembership{}
    and~Patricio~Vielva~\IEEEmembership{}%
    \thanks{D. Herranz and
    P.~Vielva are with the Instituto de F\'\i
    sica de Cantabria, CSIC-UC, Av. los Castros s/n, 39005 Santander,
    Spain, e-mail: herranz@ifca.unican.es, vielva@ifca.unican.es}%
    \thanks{Manuscript received ---; revised ---.}}


\markboth{IEEE Signal Processing Magazine, Vol.~xx,
  No.~x, Month Year}%
  {Herranz & Vielva: short title}



\maketitle

\begin{abstract}
\boldmath
We aim to present a tutorial on the detection,
parameter estimation and statistical analysis of compact sources
(far galaxies, galaxy clusters and Galactic dense emission regions)
in cosmic microwave background observations. The topic is of great relevance for current
and future cosmic microwave background missions because
the presence of compact sources in the data introduces very
significant biases in the determination of the cosmological parameters
that determine the energy contain, origin and evolution
of the universe and because compact sources themselves provide
us with important information about the large scale structure of
the universe.

\end{abstract}


% \fi
\IEEEpeerreviewmaketitle

\IEEEPARstart{C}{osmology} concerns itself with the fundamental questions about the formation, structure and evolution of the universe as a whole.
The Cosmic Microwave Background (CMB) radiation is one of the foremost pillars of Physical Cosmology.
%PATO
%(being the other two the primordial nucleosinthesis and the expansion of the universe)
%
Joint analyses
of CMB and other astronomical observations are able to determine with ever increasing precision the value of the fundamental cosmological parameters
(such as the age of the universe, its matter and energy content and its geometry) and to provide us with valuable insight about the dynamics of the universe in evolution.

The CMB radiation is a relic of the hot and dense first moments of the universe: a extraordinarily homogeneous and isotropic blackbody radiation, which however shows small temperature anisotropies that are the key for understanding the conditions of the primitive universe,
testing cosmological models and probing fundamental physics at the very dawn of time. For a short review on the CMB, see~\cite{hu02}.
%PATO
%CMB anisotropies are, actually, the seed for the formation and evolution of the large scale structure (LSS) that can be observed in the
%Universe and which is traced by galaxies and galaxy clusters (see, for instance, \cite{peacock}).
%
CMB observations are obtained by imaging of the sky at microwave wavelengths. However, the CMB signal is mixed with other astrophysical signals of both Galactic and extragalactic origin. In order to properly exploit the cosmological information contained in CMB images, they must be cleansed of these other astrophysical emissions first.

Blind source separation (BSS) in the context of CMB cosmology has been a very active field in the last few years. Most of the works have been oriented to the separation of the so-called diffuse components (sources that do not show clear borders or spatial boundaries, such as CMB itself and the cloud-like Galactic structures). A recent comparison of different BSS and other source separation techniques for diffuse sources can be found in \cite{challenge08_short}. Conversely, the term \emph{compact sources} is often used in the CMB literature referring to spatially bounded, small features in the images, such as galaxies and galaxy clusters. For reasons that we will outline in the following section, compact sources and diffuse sources are usually treated separately in CMB image processing. We will devote this tutorial to the case of compact sources. As we will see in the following sections, and particularly in Section~\ref{sec:particularities}, many of the compact source detection techniques that are widespread in most fields of astronomy are not easily applicable to CMB images. In this tutorial paper, we will make an overview of the fundamentals of compact object detection theory keeping in mind at every moment these particularities. Throughout the paper we will briefly consider Bayesian object detection, model selection, optimal linear filtering, non-linear filtering and multi-frequency detection of compact sources in CMB images.

\section{CMB and compact sources}

CMB anisotropies are extremely weak. Moreover, we cannot
observe them directly; we observe instead a mixture of CMB and other
astrophysical sources of radiation (usually referred to as \emph{foregrounds} or just \emph{contaminants}) that
lie along the line of sight of CMB photons. The foregrounds at
microwave wavelengths can have Galactic (synchrotron radiation,
\emph{brehmsstrahlung}, thermal and electromagnetic dust emission) or extragalactic
(galaxies, galaxy clusters) origin. Besides for ground-based experiments
the atmosphere must be accounted as another contaminant. In addition, obviously,
all observations are affected by instrumental noise and convolution due to the finite resolution of the
optical devices. Therefore a good source separation is an indispensable
prerequisite for any serious attempt to do CMB science. Moreover, the microwave window of the electromagnetic spectrum has been opened for observation only very recently. As a result, the microwave sky is poorly known. In particular, the present knowledge about extragalactic sources in frequencies that range from 20 to $\sim 1000$ GHz is almost completely an extrapolation from observations at lower or higher frequencies, with only a few relevant and very recent direct observations \cite{SCUBA02_short,hinshaw07_short,NEWPS07}.

Present-day CMB telescopes have relatively poor angular resolutions (from a few arcminutes to one degree). Extragalactic foregrounds show angular sizes smaller than, or at most comparable to, such scales\footnote{In astronomy, the position and size of celestial objects are described in spherical coordinates, with center in the observer. The angular size or angular scale of any object is its visual diameter measured as an angle. Throughout this paper we will use the term \emph{scale} as a synonym of angular scale.}. Therefore, they can be considered as point-like (compact) objects from the point of view of CMB experiments. The detection of faint compact objects is a very important task that is common to many branches of astronomy.

Compact sources constitute a special case among CMB
foregrounds.
%PATO
%They are the main contaminant at small angular
%scales~\cite{zotti99} and they have a dramatic impact on the determination
%of the CMB angular power spectrum (from where the determination of the cosmological
%parameters is done),
%
They are the main contaminant at small angular
scales~\cite{zotti99} and they have a dramatic impact on the estimation of
the cosmological parametersfrom the CMB signal,
on the separation of other
components and on the study of the possible non-standard cosmological scenarios (by
probing the Gaussianity and isotropy of the CMB).
Compact sources come in three main varieties: far galaxies (including
radio galaxies, quasars, blazars, 'ordinary' galaxies, dusty galaxies
with high stellar formation rate, protogalaxies, spheroids and many
other kinds of objects), galaxy clusters (leaving an imprint on CMB
radiation, mainly through inverse Compton scattering) and Galactic compact
objects (cold cores, supernova remnants, dense knots of dust,
etc). The two first classes of objects are almost uniformly
distributed across the sky, whereas the third class is mainly
concentrated near the Galactic plane. Practically all the objects
belonging to the first class have angular sizes that are much smaller
than the angular resolution of CMB experiments, while some galaxy
clusters and many Galactic objects are observed as extended
sources.
The electromagnetic spectral behavior of galaxy clusters is pretty well
known, while it is not for the other two classes of objects.
The previous sentences serve to illustrate the diversity and heterogeneity
of compact sources. Due to this heterogeneity, component separation
techniques based on generative mixing models usually fail to
achieve the separation of compact sources (the only exception to this
rule are galaxy clusters). Specific detection (identification) and
separation techniques must be tailored for compact sources.

\section{Basics of source detection in CMB images}

\subsection{Particularities of the detection problem} \label{sec:particularities}

The detection on point-like sources is a quite old problem in astronomy. Over the years, many algorithms and software codes have been developed in order to find compact objects in optical images, and these methods have been adapted, or new others have been developed, every time a new section of the electromagnetic spectrum has become available for observation. For example, the widely used \textsc{Clean} algorithm \cite{clean74} and its descendants have been at the basis of interferometric radio astronomy for the last 35 years. Other algorithms and codes such as \textsc{Daofind} \cite{DAOfind} and \textsc{SExtractor} \cite{SExtractor} have proven useful in optic, infrared and X-Ray astronomy. However, up to now none of these techniques (with the exception of \textsc{SExtractor}) have been much used in the context of compact source detection at CMB frequencies. In this section we will try to explain why it is so.

\begin{figure}[!t]
\centering
\includegraphics[width=\columnwidth]{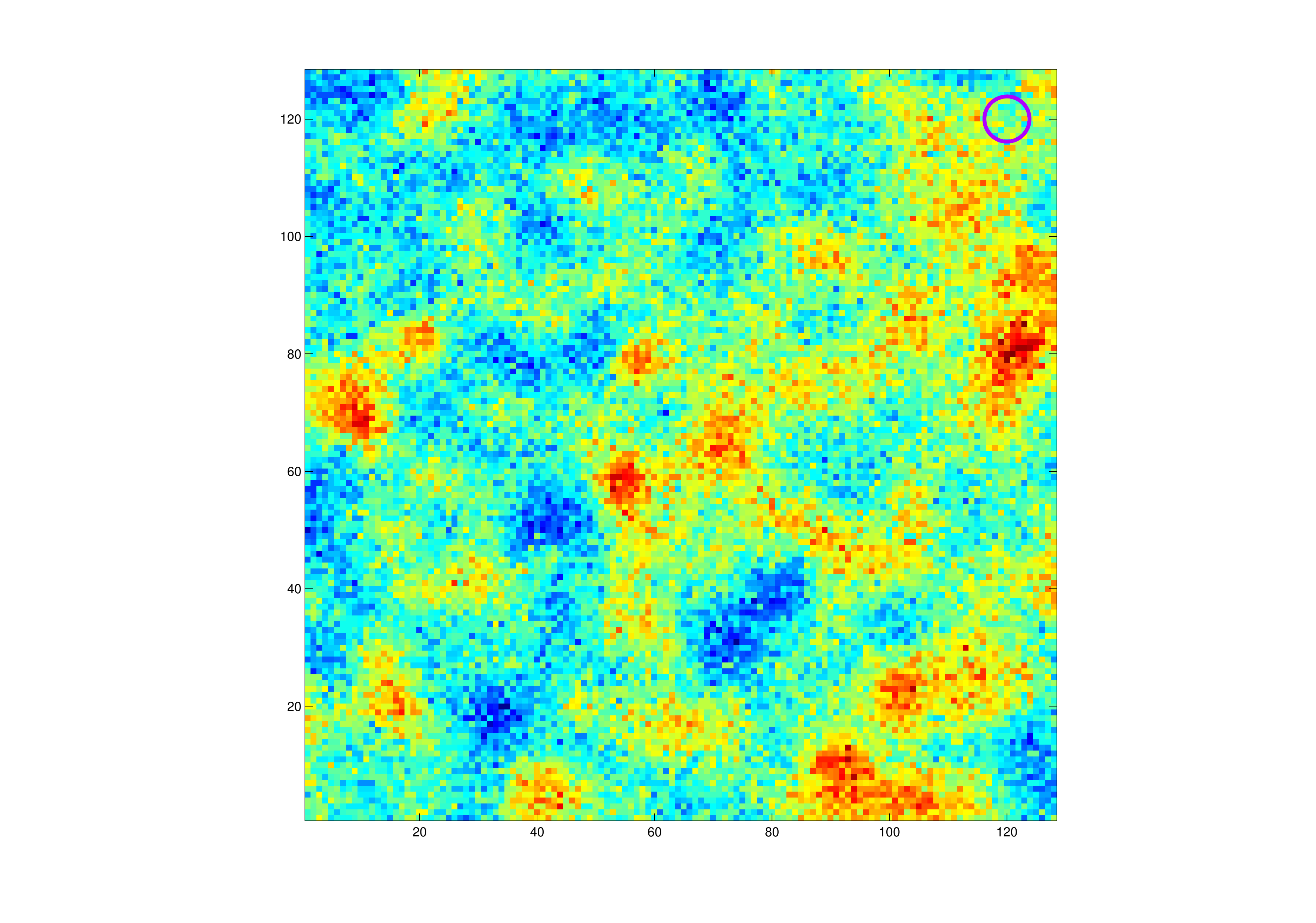}
\caption{A $14.6 \times 14.6$ square degree portion of the sky,
  centered at Galactic latitude $b=60^{\circ}$ and Galactic longitude
  $l=100^{\circ}$, observed with the WMAP satellite at 23 GHz. In the
  upper right corner we have superimposed a circle whose diameter is
  $d=0.88^{\circ}$, equal to one FWHM of the
  telescope receiver for that frequency. }
\label{fig:fig2}
\end{figure}

Figure~\ref{fig:fig2} shows a typical example of CMB image, taken
from an observation of the sky made at 23 GHz by the NASA
satellite WMAP~\cite{wmap0_short}. At first glance it is impossible to tell
apart the different signals that contribute to the image. Besides from
the detector noise, which operates at the pixel scale, all the
other components show spatial correlation. In the upper right corner
of Figure~\ref{fig:fig2} we have superimposed a circle whose
diameter is one full width at half maximum (FWHM) of the WMAP
antenna at 23 GHz: all the features in the image (apart from
instrument noise) smaller than those of the antenna are smoothed
out. Moreover, that scale is the natural one of point-like objects
such as galaxies and galaxy clusters. Note that many of the
fluctuations seen in the image have scales similar to the antenna
FWHM: as we will see, this makes it harder to detect compact
sources.

The \textsc{Clean} algorithm \cite{clean74} uses a simple iterative procedure to find the positions and strengths of these sources, going from the brightest intensity point in the image, fitting it to a source template profile, subtracting it from the data and iterating the whole process until some convergence criterion is met. In its basic implementation, \textsc{Clean} is equivalent to a least-squares minimization of the difference between the image and a model consisting of a sum of `clean' source profiles \cite{schwarz79}. \textsc{Clean} implicitly assumes that the sky can be represented by a small number of point sources in an otherwise empty field of view; unfortunately, as it is shown in Figure\ref{fig:fig2}, this is not the case for the typical CMB image, where compact sources are not dominant and easily mingle with intrinsic CMB fluctuations of similar intensity and size. Even with the recent breakthroughs in the development of fast algorithms, \textsc{Clean} methods are computationally costly and require a non negligible amount of fine tuning work in order to perform properly on CMB images.

Another critical aspect of compact source detection in CMB images is the spatial modeling of the image background. A common practice in astronomy is to subtract the image background before detecting the compact sources.
It is often assumed that the background is smoothly varying and has a characteristic
scale length much larger than the scale of the discrete objects
being sought. For example, \textsc{SExtractor} approximates the background
emission by a low-order polynomial, which is subtracted
from the image \cite{SExtractor}. Object detection is then performed by finding sets
of connected pixels above some given threshold. However, again it is not the case for CMB images such as the one shown in Figure\ref{fig:fig2}.
CMB emission  varies on
a characteristic scale of order $\leq 10$ arcminutes, comparable to the angular resolution of most CMB experiments, and cannot be safely modeled by a low order polynomial. The same applies to some Galactic components, especially the Galactic dust.

The problem, in short, can be summarized as follows: we have a unknown number of weak compact sources totally embedded in an omnipresent, non-sparse  background
%PATO
%(`noise')
%
whose intensity dominates over the compact sources and whose characteristic scale of variation is comparable to the size of the sources we are looking for. Under these circumstances, common methods such as \textsc{Clean}, \textsc{DAOfind} and \textsc{SExtractor}, however well adapted to other kind of environments may they be, find serious problems to work properly; we need to go back to the origins and review from scratch the fundamentals of source detection.

As we said previously, our ability to obtain optimal detectors depends on our degree of knowledge about the probability density functions (PDFs) of signal and noise. Notice that in the context of compact objects in CMB images, the former are referred to as the \textit{signal}, whereas the
rest of the components (i.e. the CMB itself, the Galactic contaminants and the instrumental noise) are considered as a \textit{generalized noise}.
Indeed, only the brightest of the compact objects are
actually considered as the \textit{signal}, the rest of the compact object emission, that forms a uniform background across the celestial
sphere, are also considered as a contributor to the \textit{generalized noise}.
Hereinafter we will refer to the \textit{generalized noise} just as the \textit{noise},
unless other case is explicitely said.
%PATO
%Please note that although the CMB may be Gaussian distributed, with the addition of the other Galactic and extragalactic emissions the \emph{generalized noise} is not Gaussian distributed.
%
Let us briefly describe the statistical properties of these two components for the case of the microwave sky:

\subsubsection{Compact sources}

At the low frequencies ($10-150$ GHz) they
show an almost uniform spatial distribution over the sky. Clustering
takes an increasingly important role for higher frequencies.
Regarding the flux distribution (\emph{number counts}),
the simplest models in use consider power law distributions, whose
normalization and slope are determined by fitting to observational
data. The best available models depart significantly from pure power
law descriptions and are derived from a mixing of extrapolations
from observations at other frequencies and physical modeling of the
galaxies~\cite{zotti05}.

\subsubsection{Noise}

Regarding the $n$-point PDF, standard inflationary cosmological
models predict that CMB fluctuations have a multi-normal
distribution\footnote{Some non-standard cosmological models predict
some degree of non-Gaussianity of the CMB. However, current
observational constrains imply that such degree of non-Gaussianity
must be small.}. Instrumental noise is also well modeled by a random
Gaussian process (not necessarily stationary). Galactic foregrounds,
however, are strongly non-Gaussian distributed. Moreover, the PDF of
the Galactic components is poorly known at microwave frequencies,
particularly at small angular scales and, besides, it is highly
non-stationary, depending strongly on the Galactic latitude.

\subsection{Two standpoints} \label{sec:standpoints}

Three quality indicators are considered for compact source catalogues: \emph{reliability}, \emph{completeness} and \emph{accuracy}. Please note that astronomers do not always follow the standard terminology used in statistics. In fact, different definitions appear in the literature.
%PATO
%, but confusion is avoided as long as the terms are explicitly defined for a given context.
%
We will use the following terms:
%\begin{itemize}
%  \item

\subsubsection{Reliability} is one minus the ratio of false alarms (\emph{'spurious detections'}) over the total number of alarms, for a given detection criterion:
$R  = 1 - n_e/(n_d+n_e)$,
%      \begin{equation}\label{eq:reliability}
%        R  = 1 - \frac{n_e}{n_d+n_e},
%      \end{equation}
%      \noindent
where $n_e$ is the number of false alarms (spurious detections) and $n_d$ is the number of true positives.
In some occasions, astronomers use the related term \emph{purity}, with basically the same meaning. Reliability measures how many among the positives (detections) given in a catalogue are correctly identified as such.
%  \item

\subsubsection{Completeness} is the ratio of the number of actual detections, for a given detection criterion, over the true number of objects that satisfy that criterion: $C = n_d/n_t = n_d/(n_d+n_m)$,
%      \begin{equation}\label{eq:completeness}
%        C = \frac{n_d}{n_t} = \frac{n_d}{n_d+n_m},
%      \end{equation}
%      \noindent
      where $n_m$ is the number of true objects missed by the catalogue (false negatives) and therefore $n_t$ is the total number of objects that satisfy the selection condition. For example, we say that a catalogue has a completeness of $0.99$ above 1 Jy\footnote{In radio astronomy, the flux unit or Jansky (symbol Jy) is a non-SI unit of electromagnetic flux density equivalent to $10^{-26}$ watts per square metre per hertz.} if $99\%$ of the sources whith fluxes above 1 Jy have been detected and form part of the catalogue. The completeness thus defined is equivalent to the \emph{sensitivity} of a binary classification test.
%  \item

\subsubsection{Accuracy} refers to the goodness of the estimation of the intensity --or, most commonly, flux (integrated intensity)-- of the sources.
%\end{itemize}

The importance given to each one of these quality indicators depends on the research goals at each moment. Compact sources can be a serious nuisance for the study of CMB anisotropies: from the standpoint of a pure CMB cosmologist, the most desiderable indicator of a compact source catalogue will be high completeness down to the lowest possible flux. In a second instance, flux accuracy would be highly desirable.

The opposite standpoint belongs to the astronomer solely interested in studying the number counts, spatial distribution and physical properties of objects such as galaxies and galaxy clusters. Possibly the astronomer will want to do a follow-up of the most interesting objects at other wavelengths, then a good reliability of the catalogue will be the most important quality indicator, followed by flux accuracy and, finally, completeness.

There is no easy compromise between the two standpoints. As it is well known,
there is a trade-off between reliability and completeness. Compact source detection is the art of reaching the most satisfactory point of equilibrium between these two contradicting poles.

\subsection{Enter the detection methods: ideal optimality and robustness} \label{sec:robustness}

Up to now, we have strictly spoken about statistical properties of the catalogues, that is, of the output of the detection, without referring to the detection process itself. Detecting is decision making: given a image of the CMB, that we know for sure is corrupted by noise, how many (if any) compact sources are in it, and where are they hidden? And what are their intensities?
%\footnote{Due to the large amount of far galaxies and to the poor resolution of the CMB experiments the actual probability to find a compact source in %any given image pixel is, in practice, one. Most of these sources are extremely weak and form a stationary background, known in the literature as %\emph{confusion noise}.
%In this work we focus on the detection of only the brigthest compact objects.}
To answer the first two questions, we need to choose a decision rule --the \emph{detector}-- and to somehow pre-process the image so that the efficiency of the detector is optimized. Hopefully, the way we process the image and the way we choose our detector are coordinated in order to obtain the most satisfactory results, in any of the senses described in the previous section. The study of this problems is the subject of \emph{detection theory}, which is exhaustively covered elsewhere~\cite{kay}.
In this tutorial we will focus in the reasons why compact source separation in CMB observations is a particularly difficult problem, and what aspects of detection theory are relevant for it.

The degree of difficulty of the detection problem is related to our knowledge of the signal --in this case, the compact sources-- and the noise --everything else-- characteristics in terms of their $n$-PDFs. In the ideal case when both PDFs are known and both of them are mathematically tractable it is possible, at least in theory, to obtain optimal detectors. When the PDFs are not completely known, or they are not mathematically tractable, the determination of a good enough (not \emph{optimal} any longer) detector is much more difficult. Assumptions about the signal and/or the noise PDFs are necessary to simplify the problem. But then, if the assumptions made prove to be wrong, then the outcome of the detection can be disastrous in terms of reliability, completeness, or both.  On the other hand, detectors that do not make strong assumptions about the signal and noise PDFs are further away from ideal optimality, but are less sensitive to the accuracy of our prior knowledge about the PDFs.
%
%\begin{figure}[!t]
%\centering
%\includegraphics[width=\columnwidth]{figurata.pdf}
%\caption{The trade-off between ideal optimality and robustness of a detection method.}
%\label{fig:fig1}
%\end{figure}
%
Here we find a second, empirical trade-off between ideal optimality and robustness. This trade-off has to do with
the uncertainties about the signal and the noise. Note that it is still possible to have a detection method that is at the same time ideally optimal and robust, but only if we perfectly know the statistics of both sources and noise. Again, it is necessary to reach some equilibrium between these two opposite poles.

\section{A guided tour from ideal optimality to robustness} \label{sec:tour}

As we have tried to capture in the previous section, the current
knowledge about the microwave sky is enough to give us some hints
about the PDF of the extragalactic compact sources and the noise,
but there still remain large uncertainties about both of them.
Moreover, the non-stationarity of the Galactic foregrounds makes it
very difficult to devise global detection methods valid for all
regions of the sky. Depending on how much one trusts the models
about the signal and the noise, there is a panoply of different
techniques that can be of use. In this section we will quickly
overview the different possibilities used in the literature.

Our starting point will be a complete description of the compact source detection from a Bayesian framework where all the information about the statistical properties of the sources and the background is known. Even more, Bayesian framework provides a unique paradigm for model selection. Later on, we will introduce successive simplifications,
%PATO
%used in the specialized literature,
%
of the statistical model that will allow us to relax assumptions about the background and/or the signal. These simplifications will lead to loss of ideal optimality, but in practice to gain in robustness against assumptions that may be dangerously incorrect. The first assumption we will drop is the a priori knowledge of the $n$-PDF of the sources. This will lead to Neyman-Pearson detectors. Then we will relax our assumptions about the $n$-PDF of the background and try to characterize it just in terms of its second order statistics. This will lead to the well-known matched filter. Finally, we will even drop any pretense of knowing the covariance matrix of the background: we will study tools that are able to work just with the dispersion of the 1-PDF of the background: this will lead to the use of linear band-pass filters with fixed waveform (except for an adjustable scale parameter), for example wavelet functions.

We will
focus on the case of observations made at one single wavelength; the
more complicate case of multi-wavelength detection will be discussed
in section~\ref{sec:multiw}.

\subsection{Full Bayesian approach}

Let us suppose we have a image containing a unknown number $N$ of compact sources with some given spatial templates with compact support $\tau_{i}(\mathbf{x})$, $i=1,\ldots,N$, normalized for convenience to unit peak amplitude, centered at the unknown positions $\mathbf{X}_i$ and having unknown amplitudes $A_i$, then
\begin{equation}\label{eq:model}
    d(\mathbf{x})=s(\mathbf{x})+n(\mathbf{x})=\sum_{i=1}^{N}A_i \tau_i\left( \mathbf{x}-\mathbf{X}_i  \right)+n(\mathbf{x}),
\end{equation}
\noindent
where $n(\mathbf{x})$ is a generalized noise defined as all contributions to the image aside from the discrete objects, and it is in general non-Gaussian distributed, non white and non stationary. The $i^{\mathrm{th}}$ template profile can be defined in terms of a set of parameters $\theta_i$ that include the position $\mathbf{X}_i$ of the source and some measure $R_i$ of its size. An example is the circularly symmetric Gaussian profile
\begin{equation}\label{eq:gaussian}
    \tau_G \left(\mathbf{x};\theta_i \right) = \exp\left[ -\frac{|\mathbf{x}-\mathbf{X}_i|^2}{2R_i^2}   \right].
\end{equation}
\noindent
This profile is a good approximation of point sources observed through a Gaussian beam, and in that case all $R_i=R$, the width of the beam. For the moment let us allow the sources to have different sizes. The whole set of parameters defining a single source $i$ would be then $\Theta_i=\{A_i,\theta_i\}$ and the total set of unknown parameters belonging to the compact sources in the image would be $\Theta = \{\Theta_1,\ldots,\Theta_N\}$. Bayesian inference methods provide a consistent approach to the estimation of the parameters $\Theta$ in a model $H$ for the data $\mathbf{d}$:
\begin{equation}\label{eq:bayes}
    P\left(\Theta|\mathbf{d},H   \right) = \frac{P\left( \mathbf{d} |\Theta, H \right)  P\left(\Theta | H \right)}{P\left(\mathbf{d}|H  \right)},
\end{equation}
\noindent where $P\left(\Theta|\mathbf{d},H   \right)$ is the
posterior probability distribution of the parameters, $P\left(
\mathbf{d} |\Theta, H \right)$ is the
likelihood, $P\left(\Theta | H \right)$ is the
prior and $P\left(\mathbf{d}|H  \right) \equiv E$ is the Bayesian
evidence. Inferences are usually obtained either by taking samples
from the posterior using Markov-Chain Monte Carlo (MCMC) methods or by applying Maximum A Posteriori (MAP) inference.
But for the detection problem we need to go further and to perform
model selection, that is, decide between the null hypothesis $H_0$
(`there are no sources present in the image') and an alternative
hypothesis $H_1$. This can be done by comparing their respective
posterior probabilities given the observed data $\mathbf{d}$:
\begin{equation}\label{eq:evidence_ratio}
    \frac{P(H_1|\mathbf{d})}{P(H_0|\mathbf{d})} = \frac{P(\mathbf{d}|H_1) P(H_1)}{P(\mathbf{d}|H_0) P(H_0)} = \frac{E_1}{E_0} \frac{P(H_1)}{P(H_0)},
\end{equation}
\noindent where the evidences $P(\mathbf{d}|H_1)=E_1$ and
$P(\mathbf{d}|H_0)=E_0$ can laboriously be calculated as the factor
required to normalize the posterior over $\Theta$, under the
appropriate hypothesis. Bayesian strategy can accomplish source
detection and parameter estimation in a single step. A possible
strategy consists on the following iterative algorithm:
\begin{enumerate}
  \item Define the likelihood and the prior.
  \item Locate the global maximum of the posterior using some adequate method (MCMC sampling, downhill simplex minimizer, etc). In this step it may be necessary to locally approximate the posterior distribution as a multivariate Gaussian.
  \item Estimate the parameters $\Theta_i$ corresponding to that maximum.
  \item Calculate the posterior ratio ratio (\ref{eq:evidence_ratio}) and make the decision between the null hypothesis (no detection made) or the alternative hypothesis (detection made). This step usually requires to approximate the local structure of the posterior as a multivariate Gaussian as well.
  \item If the detection has been made, subtract the source with parameters $\Theta_i$ from the data and do a new iteration from step 2.
  \item Stop the iterations when the posterior ratio criterion $P(H_1|\mathbf{d}) > P(H_0|\mathbf{d})$ fails.
\end{enumerate}
This procedure has some features reminiscent of the widely-used \textsc{Clean} algorithm \cite{clean74}, giving rise to Bayesian methods such as \textsc{McClean} and \textsc{MaxClean} \cite{hob03}.
Obviously, the crucial points of the whole process are the choice of
the likelihood and the prior (both for parameters and hypotheses),
and the way the maxima of the posterior distribution is
mapped.~\cite{hob03} has explored the problem for the case of white
Gaussian noise and a flat prior for the parameters $\Theta_i$. In a
posterior work a much faster algorithm \textsc{PowellSnakes}, based on the
simultaneous search for a number of local maxima and considering
Gaussian correlated noise in the likelihood, was developed~\cite{psnakesI}.

\subsection{The Neyman-Pearson Lemma and Generalized Likelihood Ratio Test}

The Bayesian approach is very powerful and leads to ideally optimal
results in terms of reliability and completeness of the output
catalogues if and only if the right likelihood and priors are
chosen. Unfortunately, as we saw in
section~\ref{sec:particularities}, our present knowledge of the
statistical properties of the noise (likelihood) and the signal
(prior) is somewhat less than satisfactory. Potentially dangerous
errors can come either from a bad modeling of the likelihood or from
a bad modeling of the prior. In the current state of affairs, the
uncertainties about the priors are larger than the uncertainties of
the likelihood. The reason is that the assumption that the noise is
distributed following a Gaussian with correlations is pretty good,
at least in regions far from the Galactic plane where the strongly
non-Gaussian foregrounds are subdominant. A further argument that
is usually argued for approximating the distribution by a
Gaussian grounds in The Central Limit Theorem, and is set as follows:
in most of the applications, data is analysed in terms of their
decomposition in harmonic coefficients\footnote{The harmonic
coefficients $a_{\ell m}$ of a given signal $s\left(\vec{x}\right)$
are given by
$a_{\ell m} = \int {\rm d}\Omega_{\vec{x}} s\left(\vec{x}\right) Y_{\ell m} \left(\vec{x}\right)$,
where the integral is defined over
the celestial sphere, and $Y_{\ell m} \left(\vec{x}\right)$ are the
spherical harmonic functions.} rather than in the real space; notice
that, for a given multipole $\ell$, the harmonic coefficient $a_{\ell m}$
is nothing but a linear combination of the whole observation
(weighted by the spherical harmonic functions), and, thefore
(at least for the most general scenarios) they are better
described by a Gaussian distribution, than the data in the
real space.
Following this idea, in this and the following sections we
focus on detection criteria that rely only on the noise statistics. The formalism below is valid for a general PDF of the noise, with no assumptions about Gaussianity.

Using only information about the noise statistics, the Neyman-Pearson (NP) Lemma tells us that a suitable
detection criterion is given by the Generalized Likelihood Ratio Test (GLRT):
\begin{equation}\label{eq:NP}
    \xi(\mathbf{d})=\frac{P\left( \mathbf{d} | \Theta, H_1 \right)}{P\left( \mathbf{d} | \Theta, H_0 \right)} \geq \xi_*,
\end{equation}
\noindent where $\xi_*$ is a constant which defines a region of
acceptance in the space of the data $\mathbf{d}$ and whose value is
arbitrarily set by fixing a conveniently small probability of false
detections due to the noise. Given the likelihoods $P\left(
\mathbf{d} | \Theta, H_1 \right)$ and $P\left( \mathbf{d} | \Theta,
H_0 \right)$ the test statistics defined by the GLRT (\ref{eq:NP}) has the maximum power for a fixed value of
the probability of false alarms. Two possible ways to further
increase the power of the test are to add further information by
increasing the number of observable variables and to modify $P\left(
\mathbf{d} | \Theta, H_1 \right)$ and $P\left( \mathbf{d} | \Theta,
H_0 \right)$ by means of some signal processing in order to make it
easier to discriminate between them. An example of this kind of
procedures can be seen in~\cite{can05b}, where specific linear
filters were designed in order to optimize the power of the GLRT test
for Gaussian noise in the space of variables $(\nu,\kappa)$, where
$\nu$ is the (normalized) intensity of the sources and $\kappa$
their (normalized) curvature, $\kappa \propto - \nabla^2 \tau (\mathbf{x})$ (other alternative definitions of the curvature of the sources can be used).
%PATO
%This was possible because the
%probability distributions of peaks with given values of $\nu$ and
%$\kappa$ can be obtained analytically for correlated Gaussian
%noise~\cite{rice}.
%

\subsection{Thresholding and basic linear matched filtering}

However, in many occasions it is difficult to estimate the
distribution of quantities such as the curvature of the peaks of the
noise (an example is the pure white Gaussian noise, where the
curvature of the field is not even well defined). The most frequent
situation is when one has to decide between the null and the
alternative hypotheses just on the basis of pixel intensities. In
that case, and for correlated Gaussian noise, the likelihoods in eq.
(\ref{eq:NP}) take the simple form
\begin{eqnarray}
  P\left( \mathbf{d} | \Theta, H_0 \right) & \propto & \exp \left[ -\frac{1}{2} \mathbf{d}^{T} \mathbf{C}^{-1} \mathbf{d}  \right], \nonumber \\
  P\left( \mathbf{d} | \Theta, H_1 \right) & \propto & \exp \left[ -\frac{1}{2} \mathbf{(d-s)}^{T} \mathbf{C}^{-1} \mathbf{(d-s)}  \right],
\end{eqnarray}
\noindent
where $\mathbf{C}$ is the correlation matrix of the noise, $\mathbf{s}$ is the source and the proportionality factors of both equations are both the same normalization, that is not essential for the purpose of the GLRT test. Then, it can be proven that $H_1$ has to be chosen when the statistics $T(\mathbf{d})$ is
\begin{equation}\label{eq:MF}
    T(\mathbf{d}) \propto \mathbf{d}^{T} \mathbf{C}^{-1} \mathbf{\tau} \geq \xi_*
\end{equation}
\noindent for an arbitrarily chosen threshold $\xi_*$. The threshold
is usually fixed so that the probability of having a false alarm is
conveniently small.
The proportionality factor in this equation is a normalization that can be chosen arbitrarily. A usual convention is to normalize the filter so that the amplitude of the sources is preserved.
Thresholding is the most frequently used
detection technique in astronomy. The most typically used threshold
is the so-called $5\sigma$ detection, that is, $\xi_* = 5\sigma$,
where $\sigma$ is the standard deviation of the Gaussian noise. For this case, the
statistics $T(\mathbf{d})$ is the NP detector for the problem
and the lineal operator defined by $\mathbf{C}^{-1} \mathbf{\tau}$ is
the well-known \emph{matched filter}.

\begin{figure}[!t]
\centering
\subfloat[]{
    \includegraphics[width=0.5\columnwidth]{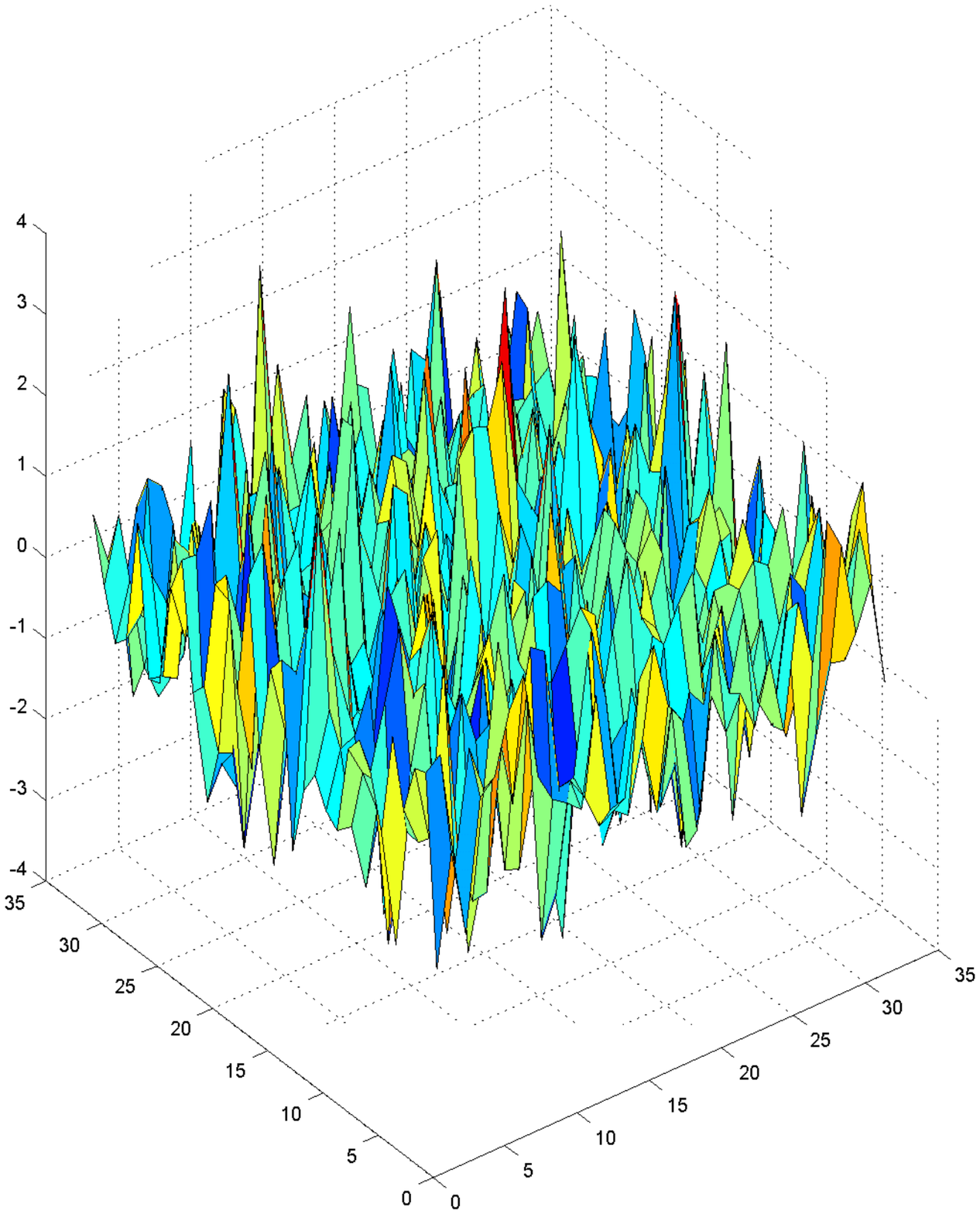}
}
\subfloat[]{
    \includegraphics[width=0.5\columnwidth]{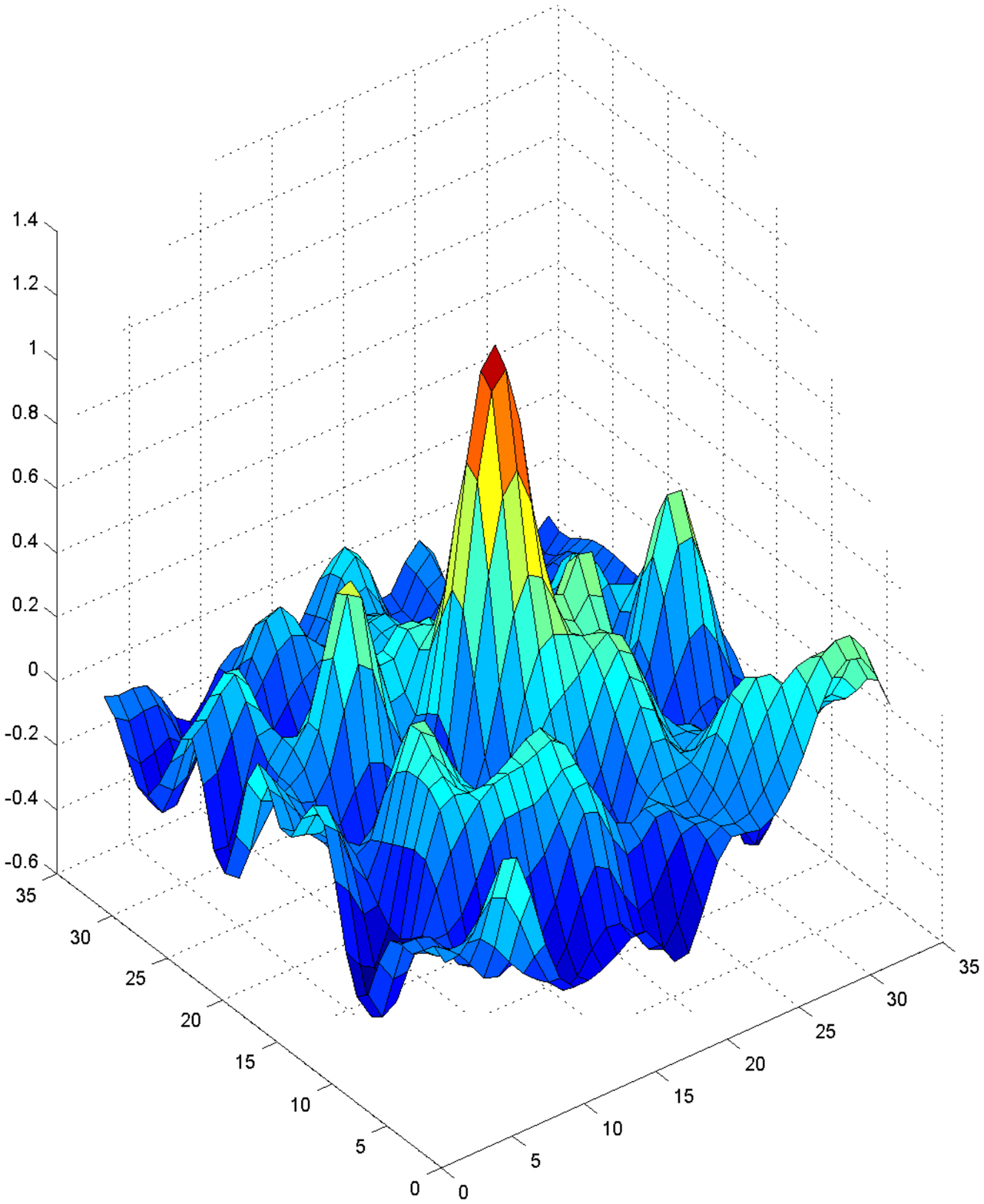}
}\\
\caption{(a) Noisy image. (b) The same image, after a matched filter has been applied.}
\label{fig:fig3}
\end{figure}

The role of the matched filter in the compact source detection is
nothing but to smooth out the noise with respect to the sources so
that we can lower the threshold for a fixed rate of false alarms.
Figure~\ref{fig:fig3} shows this effect. Panel (a) shows
a Gaussian-shaped source embedded in white noise. Visually, it is
almost impossible to detect the source. Panel (b) shows the effect
of matched filtering: now the source is clearly visible among the
noise fluctuations. Matched filtering+thresholding is generally
considered the basic default detection technique in CMB
astronomy~\cite{tegmark98}. The method can easily accommodate the
possibility of having sources with different sizes/profiles, just by
iteratively testing different $\mathbf{\tau}$ profile models and by looking for
maxima in the signal-to-noise space~\cite{herr02b}.

\subsection{Scaling linear filters: wavelets}

Whereas the matched filter is easy to implement and stands clear for
intuition, in many cases the noise correlation matrix $\mathbf{C}$
is not known. Under the assumption that sources are sparse,
$\mathbf{C}$ can be directly estimated from the data $\mathbf{d}$.
But in some cases the sampling of $\mathbf{d}$ is not good enough to
obtain a good estimation of $\mathbf{C}$. A typical example is a
image partially incomplete or covered by `bad' pixels. In other
cases, we may suspect that $\mathbf{C}$ is not a good representation
of the noise statistics but still want to use some kind of basic
thresholding detection for simplicity. For those cases, wavelets (a
kind of linear filters with a waveform that is fixed except for a
scale parameter $a$) may come in handy.

A specific wavelet basis, the \emph{Maar} or \emph{Mexican Hat}
wavelet (MHW)
has proved to be particularly efficient in many fields of astronomy,
and on the detection of compact sources at microwave frequencies in
particular. MHW thresholding is done after linear convolution of the
CMB image with the MHW kernel. The thresholding is
optimized by tuning the wavelet scale parameter $a$ such that the
convolved image presents a minimum standard deviations $\sigma$, with a normalization constrain that fixes the observed amplitude of the source after filtering. The
selection of the optimal value of the scale parameter $a$ is driven
by the statistical properties of the noise and, more particularly,
by the typical spatial variation of the noise fluctuations (see for
instance~\cite{vielva01a}).

Wavelet filtering has reached a high performance degree within the
CMB field. More efficient thresholding is obtained by considering an
additional parameter $n$, that defines the number of oscillations of
the waveform. The consideration of this extra parameter actually
provides a family of wavelet basis ---the MHW family
\cite{MHW206}---, that, in the Fourier space, reads as:
\begin{equation}\label{eq:MHWn}
\hat{\Psi}(a\mathbf{k}) \propto \frac{{a\mathbf{k}}^{2n}
e^{-{a\mathbf{k}}^2 /2}}{2^n n!},
\end{equation}
where $n=1$ corresponds to the Fourier transform of the standard MHW.
The additional degree of freedom injected by the $n$ parameter,
allows for a better discrimination between the noise
and the compact objects.

%PATO
%Once a wavelet or family of wavelets is chosen, t
%
The determination of the optimal values of the wavelet parameters
does not depend on any previous knowledge of the statistical
properties neither of the signal (the compact sources) nor the
noise: all the relevant information can be directly obtained from the data $\mathbf{d}$. Of course, different wavelets would provide different
thresholding criteria. The optimality of the MHW family comes, on
the one hand, from the fact that it is obtained as even derivatives
of the Gaussian (which, we recall, reflects the effective compact
source profile on typical CMB images) and, on the other hand, from
empirical exercises that have shown how (for any practical purpose)
a given choice of $a$ and $n$ provides a MHW with a shape very close
to the matched filter~\cite{can06}.

It is worth mentioning that, as far as we are aware, classical wavelet analysis, like multiresolution decomposition, has not been used in the context of compact source detection in CMB images. We believe the
major reason is related to a point previously commented: the generalized noise of CMB images show a lot
of structure at both the scales of the compact sources and
larger scales, and little is gained by applying
multiresolution. However, multi scale approaches using wavelets have been applied.
In particular,
%PATO
%\cite{vielva01a,vielva03}
%
\cite{vielva01a} have shown that under certain circumstances the characterization of the wavelet coefficients in a range of scales close to the optimal scale of detection helps to improve the photometry of the compact sources and the reliability of the catalogs of bright sources.

Finally, it is worth mentioning that the combination of filtering techniques such as the matched filter or the MHW with astronomical detection software such as \textsc{SExtractor} \cite{SExtractor} has very interesting possibilities. \textsc{SExtractor} gives the possibility of using a filter kernel defined by the user for the detection of compact sources, whereas photometry (integrated intensity estimation) is performed in the unfiltered map; this combined approach seems very promising \cite{vielva01a}, but a complete study of this possibility is still lacking. Note that in this case the improvement of the performance of \textsc{SExtractor} comes from the use of the MHW kernel.

\section{Multi-frequency or multi-channel detection} \label{sec:multiw}

Up to now, we have considered just the case of a single image containing a unknown number of point sources. Most microwave experiments, however, observe the sky at several different electromagnetic frequencies (channels). In that case, the additional information coming from the increased number of channels may be useful to improve the detectability of sources.

Two quite different cases must be considered here: sources whose frequency dependence is known and sources whose frequency dependence is unknown. The best example of the first case is constituted by the hot gas contained within galaxy clusters, that leaves a characteristic imprint on the CMB through inverse Compton scattering (thermal Sunyaev-Zel'dovich effect). If the frequency dependence of the compact sources is known, it is easy to incorporate it either in the Bayesian approach (see for example~\cite{hob03}), in the GLRT or in its particular case, the matched filter formalism~\cite{herr02a}. An additional problem is that many galaxy clusters have an appreciable extension in the sky and they cannot be considered as strictly point-like sources. The solution is to introduce a free scale parameter in the detector that can be adjusted in order to maximize the signal-to-noise ratio of the detected sources.

The problem is much more difficult when frequency dependence of the sources is not known. As argued in section~\ref{sec:particularities}, extragalactic sources are very heterogeneous and it is impossible to define a common frequency dependence for all the many different kinds of objects that can be observed at microwave frequencies. A possibility is to give probabilistic priors for the frequency dependence in the Bayesian framework, but the uncertainties remain so large that this approach seems dangerous at this time. Very recently a non-parametric linear filtering scheme, based in a combination of image fusion and matrices of filters that are orthogonal to the source profiles for the off-diagonal terms of the matrix, has been proposed~\cite{herranz08a}. But the field of multi-frequency point source detection remains a largely unexplored field in CMB-based cosmology.

A particular case of interest is the detection of polarized point sources. The radiation coming from the CMB, from some of the Galactic foregrounds and from the extragalactic point sources is partially linearly polarized; the study of CMB polarization opens a new window for fundamental cosmology and therefore all the previous discussion about the importance of detecting point sources serves also for the case of point source polarization. Given a particular coordinate system, linear polarization is given by the Stokes parameters $Q$ and $U$, that are not invariant quantities but depend on the orientation of the detector. On the other hand, quantity $P=\sqrt{Q^2+U^2}$ is invariant and has a physical meaning. The detection of $Q$ and $U$ can be considered a particular case of multi-channel detection, with the particularity that $Q$ and $U$ are statistically independent quantities one with respect to the other. A possibility is to use any of the previous techniques independently on the $Q$ and $U$ maps and then construct a `filtered $P$ map'. The other way around is to first construct a $P$ map and then apply some detector to it, for example a GLRT that takes into account that the statistic of the noise of the $P$ map is affected by the non-linear operations made to get it from $Q$ and $U$. Both approaches are studied in detail in~\cite{paco08}.

\section{Some recent results}

Most of the discussion we have described above has remained up to now almost purely theoretical, being the main reason that high-quality CMB measurements have become possible only very recently. The first all-sky catalogue of point sources in the frequency range from 23 to 94 GHz has been produced by the WMAP satellite~\cite{hinshaw07_short} by means of a matched filter defined in harmonic space and a $5\sigma$ threshold in the filtered space; this matched filter was defined globally for the whole sky. However, the statistical properties of the noise are highly non-uniform across the sky, due mainly to the presence of the Galactic foregrounds. A local approach, based on the MHW with index $n=2$ plus a more robust thresholding on overlapping small areas of the sky, was performed in~\cite{NEWPS07} and~\cite{massardi09}, leading to a new catalogue containing 484 radio sources whose properties in terms of reliability, completeness and flux accuracy have been determined by comparison with a complete source of bright objects obtained by ground-base observations at 20 GHz (see \cite{massardi09} for more details). These studies have shown the importance of detecting locally in small areas of the sky instead of globally. Although the statistical and physical analysis of the WMAP catalogues have given some interesting results~\cite{gnuevo08}, the number of extragalatic sources observed so far remains small.

\section{Some possible future developments}

CMB image processing is a relatively young area or research. Much work remains to be done. There is an incipient but resolved interest among CMB cosmologists to incorporate the newest ideas to solving the problem of compact sources in microwave images.

Maybe the most promising idea is related to the notions of sparsity and $l_p$-approximations. For the particular case of point-like objects the idea of sparse dictionaries comes naturally. However, the full application to CMB compact source detection has not been fully addressed. \cite{sanz_sparse} have developed a methodology that minimizes the $l_p$-norm
with a constraint on the goodness-of-fit and  have compared
different norms against the matched filter, with promising results, but their simulations are not yet realistic enough to tell for sure the performance of the method under realistic circumstances.

Another interesting possibility is to use time-frequency (or space-scale) representations to transform the data (using for example Gabor transforms or Wigner-Ville transforms) in order to better characterize the signature of compact sources as compared to the generalized noise. Again, this line of research has just begun to be opened in the context of CMB image processing \cite{herranz_TF09}.

Multiscale wavelet analysis may be useful to address another problem that is attracting the interest of the CMB community: the presence of extended compact sources of Galactic origin in the vicinity of the Galactic Plane (cold cores, supernova remnants, etc). Although works on this topic are very preliminary, we foresee that the use of multiscale techniques and compact region finders such as \textsc{SExtractor} may be a good starting point to detect and do accurate photometry of this kind of objects.

Finally, the field of non-linear filtering opens an almost infinite range of possibilities to be explored. Some of the above mentioned methods, such as \textsc{PowellSnakes} and some of the filters used for the detection of polarized point sources, include non-linear filtering in their schemes. Non-linear filtering schemes based on auto-regressive models of the background PDF have been recently introduced in the context of diffuse source separation, but the application of these ideas to compact source detection in CMB images has not been yet fully addressed. We foresee that the next few years will provide many new interesting insights, probably motivated by the analysis of new experiments such as the European satellites \emph{Planck}~\cite{planck_tauber05} and \emph{Herschel}~\cite{ATLAS0} that will allow us to detect up to several thousands of interesting objects, many of them totally new. The prospective is exciting.

\section*{Acknowledgment}

The authors thank Jos\'e Luis Sanz and Bel\'en Barreiro for useful discussions and comments. We also acknowledge partial financial support from
the Spanish Ministerio de Ciencia e Innovaci\'on project
AYA2007-68058-C03-02. PV acknowledges financial
support from the Ram\'on y Cajal programme.


\ifCLASSOPTIONcaptionsoff
  \newpage
\fi

\bibliographystyle{IEEEtran}
\bibliography{IEEEabrv,herranz_vielva}

\begin{thebibliography}{10}
\providecommand{\url}[1]{#1}
\csname url@samestyle\endcsname
\providecommand{\newblock}{\relax}
\providecommand{\bibinfo}[2]{#2}
\providecommand{\BIBentrySTDinterwordspacing}{\spaceskip=0pt\relax}
\providecommand{\BIBentryALTinterwordstretchfactor}{4}
\providecommand{\BIBentryALTinterwordspacing}{\spaceskip=\fontdimen2\font plus
\BIBentryALTinterwordstretchfactor\fontdimen3\font minus
  \fontdimen4\font\relax}
\providecommand{\BIBforeignlanguage}[2]{{%
\expandafter\ifx\csname l@#1\endcsname\relax
\typeout{** WARNING: IEEEtran.bst: No hyphenation pattern has been}%
\typeout{** loaded for the language `#1'. Using the pattern for}%
\typeout{** the default language instead.}%
\else
\language=\csname l@#1\endcsname
\fi
#2}}
\providecommand{\BIBdecl}{\relax}
\BIBdecl

\bibitem{hu02}
W.~{Hu} and S.~{Dodelson}, ``{Cosmic Microwave Background Anisotropies},''
  \emph{Annual Review of Astronomy and Astrophysics}, vol.~40, pp. 171--216,
  2002.

\bibitem{challenge08_short}
S.~M. {Leach} and {30 co-authors}, ``{Component separation methods for the
  PLANCK mission},'' \emph{\aap}, vol. 491, pp. 597--615, Nov. 2008.

\bibitem{SCUBA02_short}
S.~E. {Scott} and {14 co-authors}, ``{The SCUBA 8-mJy survey - I. Submillimetre
  maps, sources and number counts},'' \emph{\mnras}, vol. 331, pp. 817--838,
  Apr. 2002.

\bibitem{hinshaw07_short}
G.~{Hinshaw} and {21 co-authors}, ``{Three-Year Wilkinson Microwave Anisotropy
  Probe (WMAP) Observations: Temperature Analysis},'' \emph{\ApJS}, vol. 170,
  pp. 288--334, Jun. 2007.

\bibitem{NEWPS07}
M.~{L{\'o}pez-Caniego}, J.~{Gonz{\'a}lez-Nuevo}, D.~{Herranz}, M.~{Massardi},
  J.~L. {Sanz}, G.~{De Zotti}, L.~{Toffolatti}, and F.~{Arg{\"u}eso},
  ``{Nonblind Catalog of Extragalactic Point Sources from the Wilkinson
  Microwave Anisotropy Probe (WMAP) First 3 Year Survey Data},'' \emph{\ApJS},
  vol. 170, pp. 108--125, May 2007.

\bibitem{zotti99}
G.~{de Zotti}, L.~{Toffolatti}, F.~{Arg{\"u}eso}, R.~D. {Davies},
  P.~{Mazzotta}, R.~B. {Partridge}, G.~F. {Smoot}, and N.~{Vittorio}, ``{The
  Planck Surveyor Mission: Astrophysical Prospects},'' in \emph{3K cosmology},
  ser. American Institute of Physics Conference Series, L.~{Maiani},
  F.~{Melchiorri}, and N.~{Vittorio}, Eds., vol. 476, 1999, pp. 204--+.

\bibitem{clean74}
J.~A. {H{\"o}gbom}, ``{Aperture Synthesis with a Non-Regular Distribution of
  Interferometer Baselines},'' \emph{\aaps}, vol.~15, pp. 417--+, Jun. 1974.

\bibitem{DAOfind}
P.~B. {Stetson}, ``{More Experiments with DAOPHOT II and WF/PC Images},'' in
  \emph{Astronomical Data Analysis Software and Systems I}, ser. Astronomical
  Society of the Pacific Conference Series, D.~M. {Worrall}, C.~{Biemesderfer},
  and J.~{Barnes}, Eds., vol.~25, 1992, pp. 297--+.

\bibitem{SExtractor}
E.~{Bertin} and S.~{Arnouts}, ``{SExtractor: Software for source extraction.}''
  \emph{\aaps}, vol. 117, pp. 393--404, Jun. 1996.

\bibitem{wmap0_short}
C.~L. {Bennett} and {14 co-authors}, ``{The Microwave Anisotropy Probe
  Mission},'' \emph{\apj}, vol. 583, pp. 1--23, Jan. 2003.

\bibitem{schwarz79}
U.~J. {Schwarz}, ``{The Method Clean - Use, Misuse and Variations (invited
  Paper)},'' in \emph{IAU Colloq. 49: Image Formation from Coherence Functions
  in Astronomy}, ser. Astrophysics and Space Science Library, C.~{van
  Schooneveld}, Ed., vol.~76, 1979, pp. 261--+.

\bibitem{zotti05}
G.~{de Zotti}, R.~{Ricci}, D.~{Mesa}, L.~{Silva}, P.~{Mazzotta},
  L.~{Toffolatti}, and J.~{Gonz{\'a}lez-Nuevo}, ``{Predictions for
  high-frequency radio surveys of extragalactic sources},'' \emph{\aap}, vol.
  431, pp. 893--903, Mar. 2005.

\bibitem{kay}
S.~M. {Kay}, \emph{Fundamentals of Statistical Signal Processing: Detection
  Theory}, ser. Prentice Hall Signal Processing Series, A.~V. Oppenheim,
  Ed.\hskip 1em plus 0.5em minus 0.4em\relax Prentice Hall, 1998, vol.~{II}.

\bibitem{hob03}
M.~P. {Hobson} and C.~{McLachlan}, ``{A Bayesian approach to discrete object
  detection in astronomical data sets},'' \emph{\mnras}, vol. 338, pp.
  765--784, Jan. 2003.

\bibitem{psnakesI}
P.~{Carvalho}, G.~{Rocha}, and M.~P. {Hobson}, ``{A fast Bayesian approach to
  discrete object detection in astronomical data sets - PowellSnakes I},''
  \emph{\mnras}, vol. 393, pp. 681--702, Mar. 2009.

\bibitem{can05b}
M.~{L{\'o}pez-Caniego}, D.~{Herranz}, R.~B. {Barreiro}, and J.~L. {Sanz},
  ``{Filter design for the detection of compact sources based on the
  Neyman-Pearson detector},'' \emph{\mnras}, vol. 359, pp. 993--1006, May 2005.

\bibitem{tegmark98}
M.~{Tegmark} and A.~{de Oliveira-Costa}, ``{Removing Point Sources from Cosmic
  Microwave Background Maps},'' \emph{\apjl}, vol. 500, pp. L83+, Jun. 1998.

\bibitem{herr02b}
D.~{Herranz}, J.~L. {Sanz}, R.~B. {Barreiro}, and
  E.~{Mart{\'{\i}}nez-Gonz{\'a}lez}, ``{Scale-adaptive Filters for the
  Detection/Separation of Compact Sources},'' \emph{\apj}, vol. 580, pp.
  610--625, Nov. 2002.

\bibitem{vielva01a}
P.~{Vielva}, E.~{Mart{\'{\i}}nez-Gonz{\'a}lez}, L.~{Cay{\'o}n}, J.~M. {Diego},
  J.~L. {Sanz}, and L.~{Toffolatti}, ``{Predicted Planck extragalactic
  point-source catalogue},'' \emph{\mnras}, vol. 326, pp. 181--191, Sep. 2001.

\bibitem{MHW206}
J.~{Gonz{\'a}lez-Nuevo}, F.~{Arg{\"u}eso}, M.~{L{\'o}pez-Caniego},
  L.~{Toffolatti}, J.~L. {Sanz}, P.~{Vielva}, and D.~{Herranz}, ``{The Mexican
  hat wavelet family: application to point-source detection in cosmic microwave
  background maps},'' \emph{\mnras}, vol. 369, pp. 1603--1610, Jul. 2006.

\bibitem{can06}
M.~{L{\'o}pez-Caniego}, D.~{Herranz}, J.~{Gonz{\'a}lez-Nuevo}, J.~L. {Sanz},
  R.~B. {Barreiro}, P.~{Vielva}, F.~{Arg{\"u}eso}, and L.~{Toffolatti},
  ``{Comparison of filters for the detection of point sources in Planck
  simulations},'' \emph{\mnras}, vol. 370, pp. 2047--2063, Aug. 2006.

\bibitem{herr02a}
D.~{Herranz}, J.~L. {Sanz}, M.~P. {Hobson}, R.~B. {Barreiro}, J.~M. {Diego},
  E.~{Mart{\'{\i}}nez-Gonz{\'a}lez}, and A.~N. {Lasenby}, ``{Filtering
  techniques for the detection of Sunyaev-Zel'dovich clusters in multifrequency
  maps},'' \emph{\mnras}, vol. 336, pp. 1057--1068, Nov. 2002.

\bibitem{herranz08a}
D.~{Herranz} and J.~L. {Sanz}, ``{Matrix Filters for the Detection of
  Extragalactic Point Sources in Cosmic Microwave Background Images},''
  \emph{IEEE Journal of Selected Topics in Signal Processing}, vol.~5, pp.
  727--734, Oct. 2008.

\bibitem{paco08}
F.~{Arg\"ueso} and J.~L. {Sanz}, ``Filter design for the detection of compact
  sources embedded in non-stationary noise plus a deterministic background,''
  in \emph{Proceedings of the 16th European Signal Processing Conference
  (2008)}, ser. EUSIPCO 2008 Conference, Lausanne, Switzerland, Aug 2008, pp.
  1--5.

\bibitem{massardi09}
M.~{Massardi}, M.~{L{\'o}pez-Caniego}, J.~{Gonz{\'a}lez-Nuevo}, D.~{Herranz},
  G.~{de Zotti}, and J.~L. {Sanz}, ``{Blind and non-blind source detection in
  WMAP 5-yr maps},'' \emph{\mnras}, vol. 392, pp. 733--742, Jan. 2009.

\bibitem{gnuevo08}
J.~{Gonz{\'a}lez-Nuevo}, M.~{Massardi}, F.~{Arg{\"u}eso}, D.~{Herranz},
  L.~{Toffolatti}, J.~L. {Sanz}, M.~{L{\'o}pez-Caniego}, and G.~{de Zotti},
  ``{Statistical properties of extragalactic sources in the New Extragalactic
  WMAP Point Source (NEWPS) catalogue},'' \emph{\mnras}, vol. 384, pp.
  711--718, Feb. 2008.

\bibitem{sanz_sparse}
\BIBentryALTinterwordspacing
F.~{Martinelli} and J.~L. {Sanz}, ``Sparse representations versus the matched
  filter,'' in \emph{SPARS'09: Signal Processing with Adaptive Sparse
  Structured Representations}.\hskip 1em plus 0.5em minus 0.4em\relax HAL-CCSD,
  2009. [Online]. Available:
  \url{http://hal.inria.fr/docs/00/36/96/04/PDF/51.pdf}
\BIBentrySTDinterwordspacing

\bibitem{herranz_TF09}
D.~{Herranz}, J.~L. {Sanz}, and E.~E. {Kuruoglu}, ``Filtering in the
  time-frequency domain for the detection of compact objects,'' in \emph{17th
  European Signal Processing Conference (EUSIPCO 2009)}.\hskip 1em plus 0.5em
  minus 0.4em\relax Glasgow, Scotland: EUSIPCO, Aug 2009, pp. 2067--2071.

\bibitem{planck_tauber05}
J.~A. {Tauber}, ``{The Planck Mission},'' in \emph{New Cosmological Data and
  the Values of the Fundamental Parameters}, ser. IAU Symposium, A.~N.
  {Lasenby} and A.~{Wilkinson}, Eds., vol. 201, 2005, pp. 86--+.

\bibitem{ATLAS0}
S.~{Eales}, ``{The Herschel Space Observatory - uncovering the hidden
  universe},'' in \emph{37th COSPAR Scientific Assembly}, ser. COSPAR, Plenary
  Meeting, vol.~37, 2008, pp. 776--+.

\end{thebibliography}

\begin{IEEEbiographynophoto}{Diego Herranz}
received the degree in Physics from
the Universidad Complutense de Madrid, Madrid, Spain, in 1995 and the
Ph.D. degree in astrophysics from Universidad de Cantabria, Santander,
Spain, in 2002. He was a CMBNET Postdoctoral Fellow at the ISTI (CNR), Pisa,
Italy, from 2002 to 2004. He is currently at the Instituto de F\'\i
sica de Cantabria (IFCA), Santander, Spain, as UC Tenured Professor. He is
a Planck Scientist of the ESA Planck mission and Associated Researcher of the QUIJOTE experiment. His research
interests are in the areas of cosmic microwave background astronomy
and extragalactic point source statistics as well as the application
of statistical signal processing to astronomical data.
\end{IEEEbiographynophoto}

\begin{IEEEbiographynophoto}{Patricio Vielva}
received the degree in Physics in 1998 and the Ph.D. degree in 2003,
both from the Universidad de Cantabria (UC), Santander, Spain.
In 2004, he was a Postdoctoral Researcher at the Coll\`ege de
France/APC, Paris, France; and a Visiting Researcher at the
Cavendish Laboratory of the University of
Cambridge, Cambridge, UK, during 2006.  He is currently a \emph{Ram\'on y Cajal}
Researcher at IFCA.  His field of research is cosmology,
especially the
development and application of statistical and signal
processing tools for the
study of the cosmic microwave background and the
large scale structure of the universe.
He is a Planck Scientist of the ESA Planck mission and an Associated Researcher of the
QUIJOTE experiment.
\end{IEEEbiographynophoto}


%\newpage



%\vfill


\end{document}